\documentclass[aps,pra,floats,floatfix,twocolumn,longbibliography]{revtex4-1}

\usepackage{color} 
\usepackage{latexsym}
\usepackage{amsmath} 
\usepackage{amssymb} 
\usepackage{bm}
\usepackage{wasysym}

\usepackage{graphicx}
\usepackage{epstopdf}
\graphicspath{{./}{./Figs/}}

\usepackage[colorlinks,linkcolor=blue,urlcolor=blue,bookmarks=true,pdftitle={}]{hyperref}

\newcommand{\ket}[1]{\left| #1 \right\rangle}
\newcommand{\braket}[1]{\left\langle #1 \right\rangle }

\newcommand{\Braket}[2]{\left\langle #1 \middle| #2 \right\rangle}
\newcommand{\BraKet}[3]{\left\langle #1 \middle| #2 \middle| #3 \right\rangle}

\newcommand{\mylabel}[1]{\label{#1}} 
\newcommand{\beq}{\begin{eqnarray}}
\newcommand{\eeq}{\end{eqnarray}} 
\newcommand{\be}[1]{\begin{eqnarray}\ifthenelse{#1=-1}
{\nonumber}{\ifthenelse{#1=0}{}{\mylabel{e#1}}}}
\newcommand{\ee}{\end{eqnarray}} 

\newcommand{\hide}[1]{}
\newcommand{\Eq}[1]{\textcolor{blue}{Eq.\!\!~(\ref{#1})}} 
\newcommand{\Fig}[1]{\textcolor{blue}{Fig.}\!\!~\ref{#1}}

\begin{document}

\title{\textbf{Chaos induced breakdown of Bose-Hubbard modeling}}

\author{Sayak Ray$^1$, Doron Cohen$^2$, Amichay Vardi$^1$}

\affiliation{
\mbox{Department of Chemistry, Ben-Gurion University of the Negev, Beer-Sheva 84105, Israel} 
\mbox{Department of Physics, Ben-Gurion University of the Negev, Beer-Sheva 84105, Israel} 
}

\begin{abstract}
We show that the Bose-Hubbard approximation fails due to the emergence of chaos, even when excited modes are far detuned and the standard validity condition is satisfied. This is formally identical to the  Melnikov-Arnold analysis of the stochastic pump model. Previous numerical observations of Bose-Hubbard breakdown are precisely reproduced by our simple model and can be attributed to many body enhancement of chaos.
\end{abstract}

\maketitle

\section{Introduction}

The Bose-Hubbard model (BHM) \cite{Fisher89,Leggett01} is one of the most prominent tools in the study of interacting many-body systems, no less significant than its celebrated fermionic counterpart \cite{Hubbard63,Gutzwiller63}.  Its use for treating ultracold bosons in shallow optical lattices \cite{Bloch05,Krutitsky16}, resulted in landmark experimental demonstrations of  the superfluid to Mott insulator  quantum phase transition \cite{Jaksch98,Orzel01,Greiner02a} and of dynamical quantum phase revivals \cite{Greiner02b,Will10}. These experiments and the subsequent development of quantum engineering techniques provide an unique opportunity to simulate various models and to explore the exotic phases of quantum matter using ultracold atoms \cite{Lewenstein07,Bloch12}.

The simplest BHM includes only two modes. Originally introduced in nuclear physics by Lipkin, Meshkov, and Glick \cite{LMG}, two-mode theories apply equally well to Bose-Einstein condensates (BECs) of atoms in the same hyperfine state confined in a double-well trap, and to spinor BECs of atoms with two hyperfine states confined in a single-well trap. Two-mode BHMs are used to describe schemes for squeezing and entanglement \cite{Jo07,Esteve08,Riedel10}, Josephson oscillations and self-trapping in bosonic Josephson junctions \cite{Milburn97,Smerzi97,Albiez05,Gati07,Levy07}, the dynamical growth of quantum fluctuations \cite{Vardi01,Anglin01,Khodorkovsky08,Boukobza09,Boukobza10,Chuchem10}, and persistent currents and phase slips in superfluid atom circuits \cite{Wright13,Eckel14,Fialko12,Arwas16,Arwas17a,Arwas17b}.

%
Efforts towards realization of quantum simulators assume the validity of the BHM for describing cold atoms in optical lattices. The determination of validity criteria for the BHM  is therefore of utmost importance. While Bose-Hubbard dynamics can be affected by coupling to the ever present excited Bloch bands \cite{Cederbaum,Bergeman06,Trujillo-Martinez09,Smerzi11,Modugno18,Cataldo13}, it is generally accepted that the motion can be restricted to the lowest band of the site-chain, provided that the interaction energy is small with respect to the gap between it and the first excited band. \cite{Leggett01,Gati07,Milburn97,Anglin01}.

%
Despite this convention, the BHM was numerically shown to fail {\em within} its expected validity regime \cite{Cederbaum}. Comparing multiconfigurational time-dependent Hartree for bosons (MCTDHB) simulations of the exact dynamics, with two-mode Bose-Hubbard dynamics, deviations from the BHM were observed even for tight traps where the standard validity condition is satisfied. The cause of this surprising failure remained unclear. No mechanism was offered to explain it or to attribute it to a particular dynamical feature, and no revised validity condition has been suggested.

%
Here, we show that even if its standard validity condition is satisfied, 
the BHM is still prone to failure due to the emergence of dynamical chaos.
The naive assumption that high lying orbitals that do not participate in the dynamics 
merely renormalize the hopping elements via virtual transitions does not generally apply. 
Rather, chaos can be induced via the Melnikov-Arnold mechanism, as in Chirikov's stochastic pump model \cite{Chirikov}.
Specifically,  near-separatrix dynamics becomes stochastic due to the coupling with a high lying orbital. 
Beyond this zero order resonance, higher order resonances also show up and affect the dynamics. 
Consequently the lower band becomes entangled with higher bands, 
as reflected in reduced subsystem entropy measures.      
The previous observations of BHM breakdown \cite{Cederbaum}
can thus be attributed to emergence of chaos and generation of entanglement, and are fully reproduced by a simple three mode BHM.

\section{The BHM}
\subsection{Validity criteria}
The experimental parameters of a linear 1D chain are the axial trap frequency $\omega_{\parallel}$, the barrier transmission coefficient $T$, and the atom number $N$. The 1D interaction strength is $\lambda_0=2\hbar \omega_\perp a_s$, 
where $a_s$ is the $s$-wave scattering length, and $\omega_\perp$ is the transverse trap frequency.  The atoms mass is $m$. 
These parameters define three characteristic length-scales: 
the axial trap size $L=\sqrt{\hbar/m \omega_{\parallel}}$, 
the healing length $l_c=\sqrt{\hbar/2m\lambda_0 n}$, 
and the mean distance between atoms ${d=1/n}$, 
where ${n=N/(2L)}$ is the average atom density.
The standard BHM validity criterion assumes that the interaction energy is 
too small to bridge the $\Omega=\hbar\omega_{\parallel}$ gap between the lowest Bloch band and the
first excited band. In terms of characteristic lengths, this means $\nu \ll 1$, where ${\nu \equiv(L/l_c)^2 = \lambda_0 n/(\hbar\omega_{\parallel})}$.   
The $M$-site chain is then described by the tight-binding  Hamiltonian,
\begin{equation}
\hat{H}_{\text{BH}} = 
\frac{U}{2} \sum_{i=1}^{M} \hat{n}_i(\hat{n}_i-1)
-\frac{K}{2}\sum_{\langle ij \rangle}^M(\hat{a}_i^{\dagger}\hat{a}_j + \hat{a}_j^{\dagger}\hat{a}_i)
\label{BHH}
\end{equation}
where $\hat{a}_i$ and $\hat{n}_i$ are the bosonic annihilation and number operators,
$\langle ij \rangle$ denotes summation over nearest neighbors, 
and the effective parameters are the tunnel-splitting $K \approx \hbar\omega\sqrt{T}$,  
and the interaction-strength $UN \approx \lambda_0 n$. The standard BHM validity criterion ${\nu \ll 1}$ then takes the form,
\begin{equation} 
u \ll \frac{\Omega}{K}~,~u \equiv \frac{NU}{K}.
\label{eStandard}
\end{equation} 
where $u$ is the dimensionless interaction parameter.

\subsection{Unexpected breakdown}
In Ref.~\cite{Cederbaum}, strong deviations from the BHM were observed for $u~\sim 2$, even when $\nu=1/14$, leading to the conclusion: ``Clearly, we have shown a failure of GP theory and the BH model within their range of expected validity". Beyond the empirical determination that this unforeseen failure is ``associated with fragmentation and correlations not captured by the standard theories", no concrete mechanism has been suggested so far, for its origin.

Below, we establish that the underlying mechanism for the observed failure of the BHM is related to the robustness of chaos in the vicinity of separatrix regions. In order to investigate the chaos-induced breakdown of the BHM, we consider the $M=2$ (dimer) case, and emulate the effect of excited Bloch bands by adding to \Eq{BHH} a single detuned bosonic mode:
\begin{equation}
\hat{\mathcal{H}} = \hat{H}_{\text{BH}} + \Omega \hat{n}_0 
+ \frac{U}{2}\hat{n}_0(\hat{n}_0-1) 
- \frac{\kappa}{2} \sum_{i=1}^{2}(\hat{a}_i^{\dagger}\hat{a}_0 + \text{h.c.})
\label{BHM_trimer}
\end{equation} 
where $\Omega$ and $\kappa$ are respectively, the detuning and coupling of the auxiliary bosonic mode `0'. 
For an actual double well potential the above parameters including $\Omega$ and $\kappa$ can be estimated from the first principles as explaiend in Appendix\ \ref{A1_def}.
In our calculation we set $\hbar = 1$ and determine the units of time such that the hopping frequency is ${K=1}$. 

\section{Dynamics}

\subsection{Quantum, classical, and semiclassical propagation}
The motional constants of the Hamiltonian \Eq{BHM_trimer} are the total three-mode energy $E$ and the total three-mode particle number $N$. In what follows, we distinguish between three types of propagation: 
(a) {\em Quantum dynamics} refers to the full many-body dynamics, i.e. the time-evolved many-body state is computed as $|\Psi(t)\rangle=e^{-i \hat{\mathcal{H}} t}|\Psi(0)\rangle$, for a given initial state $|\Psi(0)\rangle$; 
(b) {\em Classical dynamics} refers to the large $N$ Gross-Pitaevskii mean-field theory, in which the field operators $\hat{a}_i$ are replaced by classical $c$-numbers $a_i=\sqrt{n_i} \exp(i\phi_i)$. Using $N$ conservation to eliminate the overall phase, the classical canonical variables are the dimer's population imbalance ${n = n_1 - n_2}$, the relative phase ${\phi = \phi_1 - \phi_2}$, 
and the auxiliary mode's population $n_0$ and phase $\phi_0$; 
(c) {\em Semiclassical dynamics} refers to the truncated Wigner classical propagation of a cloud of classical points that emulates the phase space distribution of the quantum state and averaging it to obtain the pertinent obsevables. 
For further detail on the dynamical equations, see Appendix\ \ref{A1}.

\begin{figure}[t]
\centering
\includegraphics[clip=true,width =1\columnwidth]{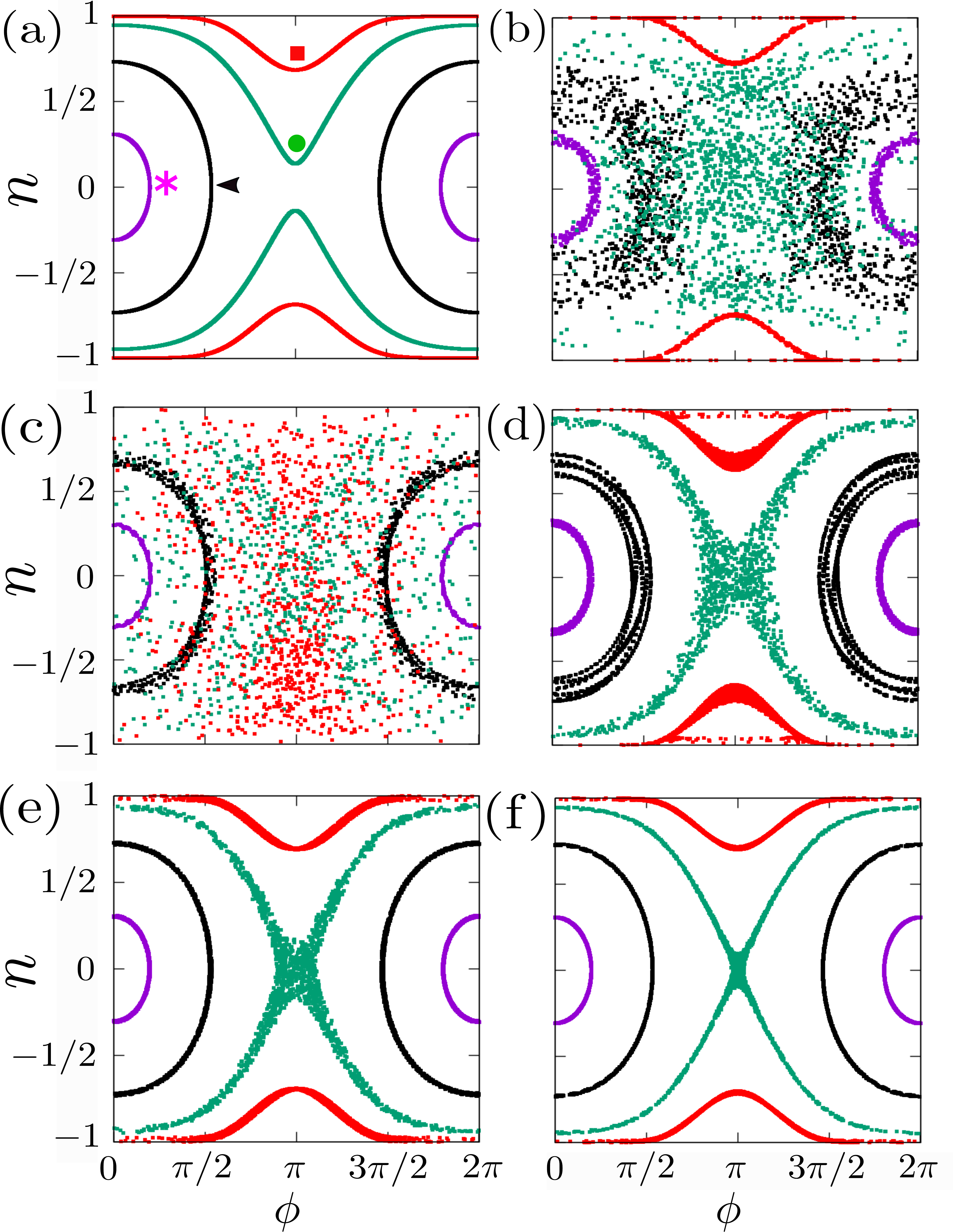}
\caption{(Color online) 
Classical Bose-Hubbard dynamics. 
Panel (a) is for an isolated two modes system. 
It shows representative Rabi-Josephson (magenta-star, black-arrow), 
near-separatrix (green-circle), 
and self-trapped (red-square) trajectories for ${u = 3}$  and ${\kappa=0}$. 
Panels (a-f) are for a two-mode system coupled to a third detuned mode (${\kappa=0.5}$), 
with $\Omega=0.5,2.0,4.5,6,7$, respectively.
They show  $\phi_0=0$ Poincare sections for the same initial conditions as in (a), 
while the third orbital is initially empty (${n_0=0}$).}
\label{pp_cl}
\end{figure}

\subsection{Semiclassical perspective}
In \Fig{pp_cl}(a), we plot the pendulum-like classical phase space of the dimer  \cite{Albiez05,Gati07,Chuchem10}, where $\kappa=0$. Since the isolated dimer has just one degree of freedom, its motion is necessarily integrable and in fact solvable  by the algebraic Bethe ansatz \cite{Guan03}. The dimensionless interaction parameter $u$ distinguishes between three {\em interaction regimes} \cite{Leggett01,Gati07,Chuchem10}: Rabi ($u{<}1$); Josephson ($1{<}u{<}N^2$); and Fock ($u{>}N^2$). 
Within the Josephson interaction regime, the phase-space consists of a low energy Rabi-Josephson oscillation region \cite{Milburn97,Levy07} and a high-energy self-trapping region \cite{Smerzi97,Albiez05,Gati07}, that are separated by a mid-energy separatrix \cite{Chuchem10}.

In panels \Fig{pp_cl}(b)-(f), we set the coupling to $\kappa=0.5$ and study the effect of the third mode for various values of its detuning $\Omega$. It should be noticed that different trajectories do not have the same energy~$E$, hence they do not belong to the same Poincare section. The addition of a third mode opens the way to non-integrable motion \cite{Korsch09,Hiller09,Liu07,Jung06,Penna03,Tikhonenkov13,Gallemi15,Dey18,Burkle19} resulting in stochastic regions in phase space due to non-linear resonances. Trajectories within these stochastic regions exhibit sensitive dependence on initial conditions, that can be quantified in terms of Lyapunov exponents. However, below we do not focus on the quantification of this sensitivity, but rather on the consequences of the chaotic motion.

Consider first the low $\Omega$ panels \Fig{pp_cl}(b)-(c). In this regime \Eq{eStandard} is not satisfied, and the trajectories are strongly affected by the coupling to the excited mode. Not only the separatrix motion is affected: in panel~(b) most Rabi-Josephson trajectories become chaotic; while in panel~(c) chaotic motion is obtained mainly for self-trapped trajectories.

By contrast, for the large $\Omega$ panels \Fig{pp_cl}(d)-(f), the validity criterion of \Eq{eStandard} is satisfied, and one expects the far-detuned mode to have a negligible effect. Indeed, this seems to be the case in most regions of phase space. However, a stochastic strip remains in the vicinity of the separatrix, even when the detuning is large. This is expected from the standard theory of non-linear resonances, and can be treated using the same procedure used for the Melnikov-Arnold analysis of the stochastic pump model \cite{Chirikov}. Thus, due to chaos, the BHM breaks down for near-separatrix motion, and in the vicinity of resonances, {\em irrespective of the  standard validity condition}.

\begin{figure}[t]
\centering
\includegraphics[clip=true,width =\columnwidth]{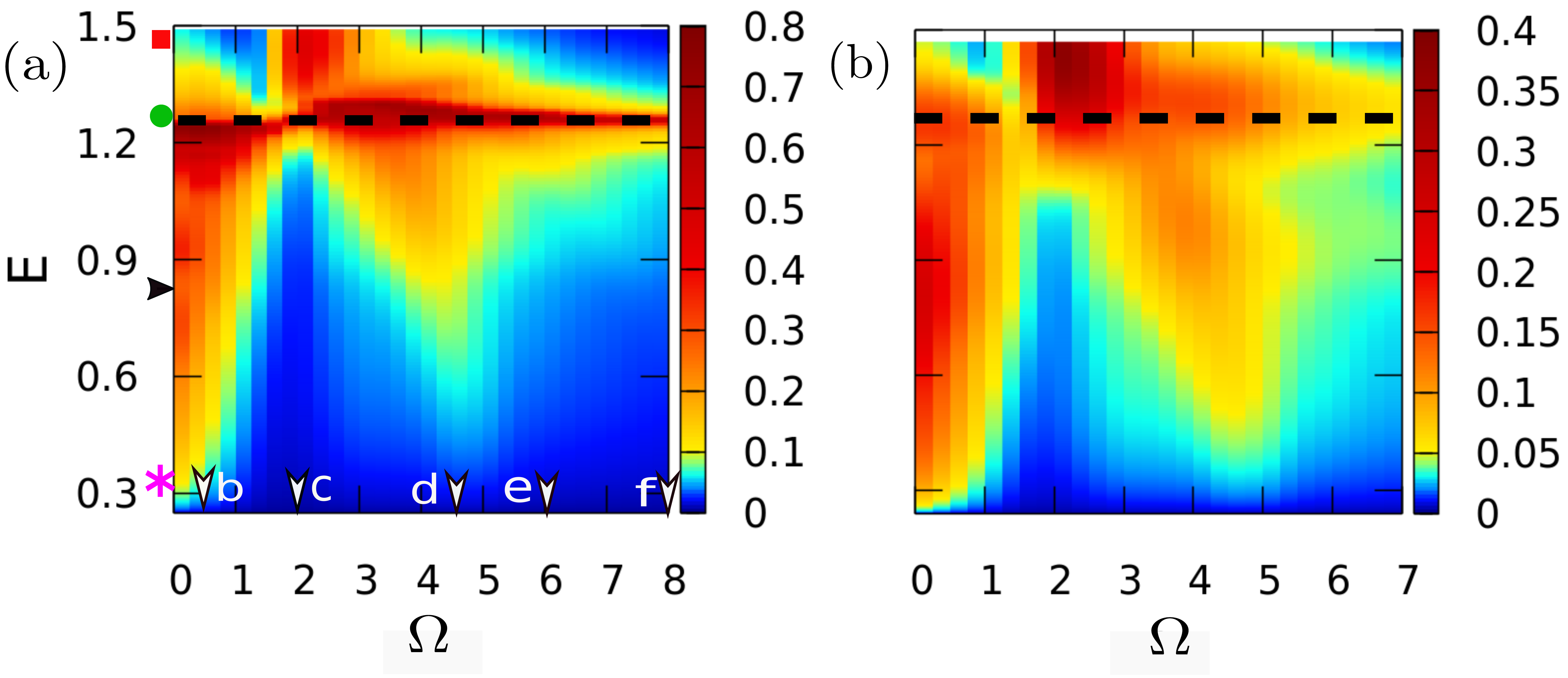}
\caption{(Color online) The deviation $d$ of \Eq{dev}, plotted as a function of the trajectory's energy $E$ and the detuning $\Omega$: (a) classical simulations; (b) quantum simulations with $N=100$, launched at the corresponding coherent state. The parameters $u=3$ and $\kappa=0.5$ are the same as in \Fig{pp_cl}. Dashed line marks the separatrix energy. The detuning values of \Fig{pp_cl}(b)-(f) are marked on the horizontal axis, and the energies of the plotted trajectories are marked on the vertical axis with the same marker-convention. Here and throughout the manuscript, we use dimensionless energy and time scales, e.g. $\Omega\rightarrow\Omega/K$, $E\rightarrow E/K$, $t\rightarrow Kt$.}
\label{f2:dev}
\end{figure}

\subsection{Deviation measure}
The deviation from the BHM is quantified by the instantaneous difference between the population imbalance $n(t)$ obtained by including the extra mode, and the same quantity $n_{\mathrm{BH}}(t)$ obtained in the two-mode approximation, averaged over $T = 2\pi$,
\begin{equation}
d(E,\Omega) = \frac{1}{T}\int_0^{T} |n(t) - n_{\mathrm{BH}}(t)| dt
\label{dev}
\end{equation}
In \Fig{f2:dev}(a) we plot $d$ as function of the trajectory's energy $E$ and the detuning $\Omega$. Since all simulations are launched with $n_0(0)=0$, the total energy $E$ equals the initial dimer energy.
One observes that the deviation $d$ is large for trajectories in regions of stochastic motion. 
The corresponding quantum results are displayed in \Fig{f2:dev}(b). 
Initial states $|\Psi(0)\rangle$ in these quantum simulations were three-mode coherent states $|n,\phi,n_0,\phi_0\rangle$,  
where $n,\phi,n_0,\phi_0$ are the same as the initial values of the classical parameters (see Appendix\ \ref{A1}). We observe good quantum-classical agreement, with some blurring of classical features due to the finite uncertainty width of the initial coherent state.

\begin{figure}
\centering
\includegraphics[width =\columnwidth]{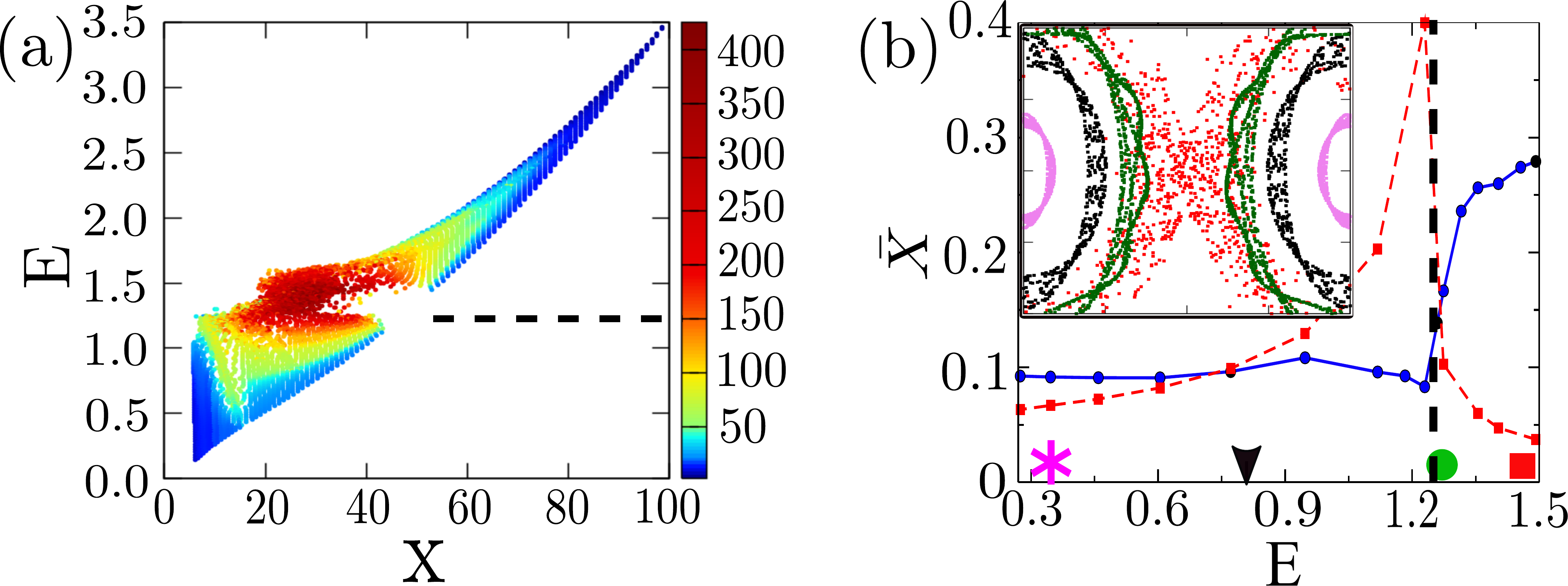}
\caption{ (color online) 
(a) The eigen-spectrum $\{ E_{\nu} \}$ obtained from diagonalization of the Hamiltonian \Eq{BHM_trimer} for $N = 100$, $u=3$, $\kappa = 0.5$, $\Omega = 2$.
Each point is positioned horizontally according to $X_{\nu}=\braket{\hat{n}_0}$,  
and color coded by participation number $M_{\nu}$. Strong mixing is witnessed at the separatrix energy (dashed line) and extends to the region above it.    
(b) Time averaged occupation of the third mode, $\bar{X} \equiv \overline{n_0(t)}$, 
obtained from three-mode classical dynamics (solid blue) as a function of $E$, 
compared with the estimate $\bar{X} \sim \kappa^2/\omega^2(E)$ (red dashed) 
that assumes quasi-integrable orbits. The separatrix energy is indicated by vertical dashed line,   
while the energies of the trajectories in \Fig{pp_cl} are indicated by symbols. 
The inset shows the stroboscopic map for the two mode BHM in the presence of driving with frequency $\Omega=2$ and intensity $A=\sqrt{\bar{X}}$. It corresponds to \Fig{pp_cl}c and has same axes.}
\label{fig:spectrum}
\end{figure}

\section{Many body enhancement of chaos} 
The emergence of stochastic regions in \Fig{pp_cl} and the deviation depicted in \Fig{f2:dev} would have been obtained also if the two-mode system is driven at frequency $\Omega$. However, as shown below, the many-body aspect of incorporating an auxiliary mode in the Hamiltonian, amounts to considerable enhancement of chaos with respect to the corresponding driven system.

Consider the representative quantum spectrum in \Fig{fig:spectrum}(a). Each point represents an exact eigenstate $\nu$ of the Hamiltonian in \Eq{BHM_trimer}, positioned horizontally according to $X_{\nu}=\BraKet{\nu}{\hat{n}_0}{\nu}$, vertically according to its energy $E_{\nu}$, and color coded by its participation number $M_{\nu} = \left(  \sum_m |\Braket{m}{\nu}|^4  \right)^{-1}$, where $m$ labels the unperturbed eigenstates of the uncoupled (${\kappa=0}$) system. Large $M_{\nu}$ implies that many eigenstates are mixed due to the ${\kappa \ne 0}$ coupling.

The Floquet states of any driving scenario would only mix $\ket{m}$ states with the same $n_0$, i.e. driving corresponds to {\em vertical} mixing in \Fig{fig:spectrum}(a). By contrast, referring to the ${(X,E)}$ diagram of the three-mode spectrum, we see that chaos in the vertical direction induces mixing also in the horizontal direction. Thus, the possibility of back action by the two mode dynamics on the auxiliary mode, results in enhanced chaos.

Considering the form of the two-mode dynamical equations ${\dot{a}_j = ... + i(\kappa/2)a_0}$ (see Appendix~\ref{A1}), the corresponding c-number driving would be obtained by  substituting ${a_0 = \sqrt{A} \exp(-i\Omega t)}$, where $A$ is a free constant parameter. To reproduce the effect of the third mode, the effective drive intensity $A$ should reflect its occupation~$n_0$. Since from inspection of \Fig{fig:spectrum}(a) it is clear that this occupation is larger in regions of chaos due to the many body mixing, the third mode's effect corresponds to an amplified drive intensity in these regions, i.e. to {\em many body enhancement of chaos}. This observation is somewhat reminiscenct of the {\em dynamical enhancement of small perturbations} in the nuclear physics context \cite{Flambaum}.
 
We may attempt to evaluate the effective $A$ from the time averaged occupation in the 3rd orbital $\bar{X}=\overline{n_0(t)}$. If chaos was not present, we could estimate $\bar{X}$ from the equation of motion ${\dot{a}_0 \approx i(\kappa/2)(a_1+a_2)}$ leading to ${\bar{X} \sim \kappa^2/\omega^2(E)}$, where $\omega(E)$ is the frequency of the unperturbed dimer oscillations. The comparison of this estimate with the actual three-mode result in  \Fig{fig:spectrum}(b), demonstrates the enhancement of $\bar{X}$ in chaotic regions. The inset in \Fig{fig:spectrum}(b) shows the dynamics of a driving scenario where $A=\bar{X}$ is the numerically extracted actual value of $\bar{X}$. Comparing with  \Fig{pp_cl}c we realize that even this procedure still underestimates the enhanced chaos.           

\begin{figure}[t]
\centering
\includegraphics[width =\columnwidth]{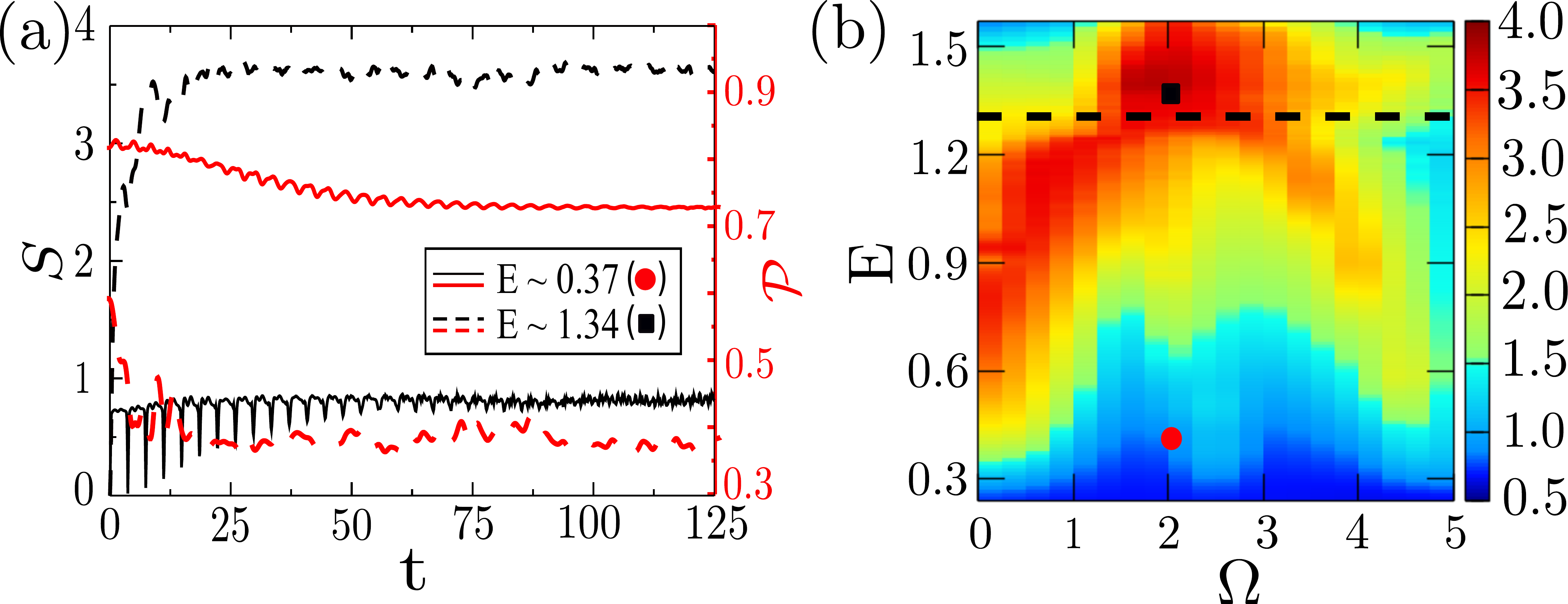}
\caption{ (color online) 
(a) Time evolution of the entanglement entropy $S$ 
and the single-particle purity $\mathcal{P}$ 
for a chaotic (dashed) and for a quasi-integrable (solid) 
three-mode dynamics, starting from an eigenstates of the unperturbed two-mode system. 
(b) The time-averaged entanglement entropy $\overline{S(t)}$ as a function of $E$ and $\Omega$. 
Markers indicate the parameter values for the curves in (a). 
The other parameters are as in \Fig{f2:dev}.}
\label{fig:entanglement}
\end{figure} 

\section{Fragmentation and entanglement} 
\subsection{One particle purity} 
In \Fig{fig:entanglement}a we prepare the system in representative eigenstates $|m\rangle$, with vacant 3rd orbital ($n_0=0$), of the unperturbed (${\kappa=0}$) Hamiltonian \Eq{BHM_trimer}, and plot the time dependence of the one-particle purity $\mathcal{P} ={\mathrm Tr}\left(\rho_{\mathrm sp}^2\right)$, where $\rho_{\mathrm sp}=\langle {\hat a}_i^\dag {\hat a}_j \rangle$ is the reduced single particle density matrix. The one-particle purity value lies in the range  $1/3 < \mathcal{P} < 1$ and its inverse indicates the number of modes required to capture the dynamics. Thus $\mathcal{P}\approx 1$ indicates the validity of mean-field theory, wherein all particles occupy a single orbital, $1/2 < \mathcal{P} < 1$ indicates two-orbital dynamics (strictly speaking, this range will also be obtained if the pertinent two orbitals project onto the auxiliary mode), while $\mathcal{P} < 1/2$ clearly indicates the breakdown of the two-mode approximation. The latter is observed if the dynamics is affected by chaos. 

\subsection{Entanglement entropy}  
The addition of an excited mode also implies that {\em entanglement} could be generated between the dimer modes and the high frequency mode. The entanglement entropy is defined as $S=\text{Tr} (\rho_{\mathrm d} \ln \rho_{\mathrm d})=\text{Tr} (\rho_0 \ln \rho_0)$, where $\rho_{\mathrm d}=\text{Tr}_0 (\rho)$ and $\rho_0=\text{Tr}_{\mathrm d} (\rho)$ are the many-body reduced density matrices of the dimer and the auxiliary mode, respectively.
In \Fig{fig:entanglement}(a) we plot the time evolution of $S$ for a couple of representative simulations. 
Carrying out such simulations at various values of $\Omega$,  for all dimer eigenstate preparations (distinguished by $E$) we plot the time-averaged $\overline{S(t)}$  in \Fig{fig:entanglement}(b). 
We observe that the entanglement entropy is large in region that support chaotic motion. Comparing with \Fig{f2:dev} we see that chaos dominates at the vicinity of the separtrix: just below it for ${\Omega<1.5}$, and just above it for ${\Omega>1.5}$. Corresponding red regions below and above the separatrix energy in \Fig{fig:entanglement}(b) show the entanglement fingerprint of chaos.

\begin{figure}[t]
\centering
\includegraphics[clip=true,width =\columnwidth]{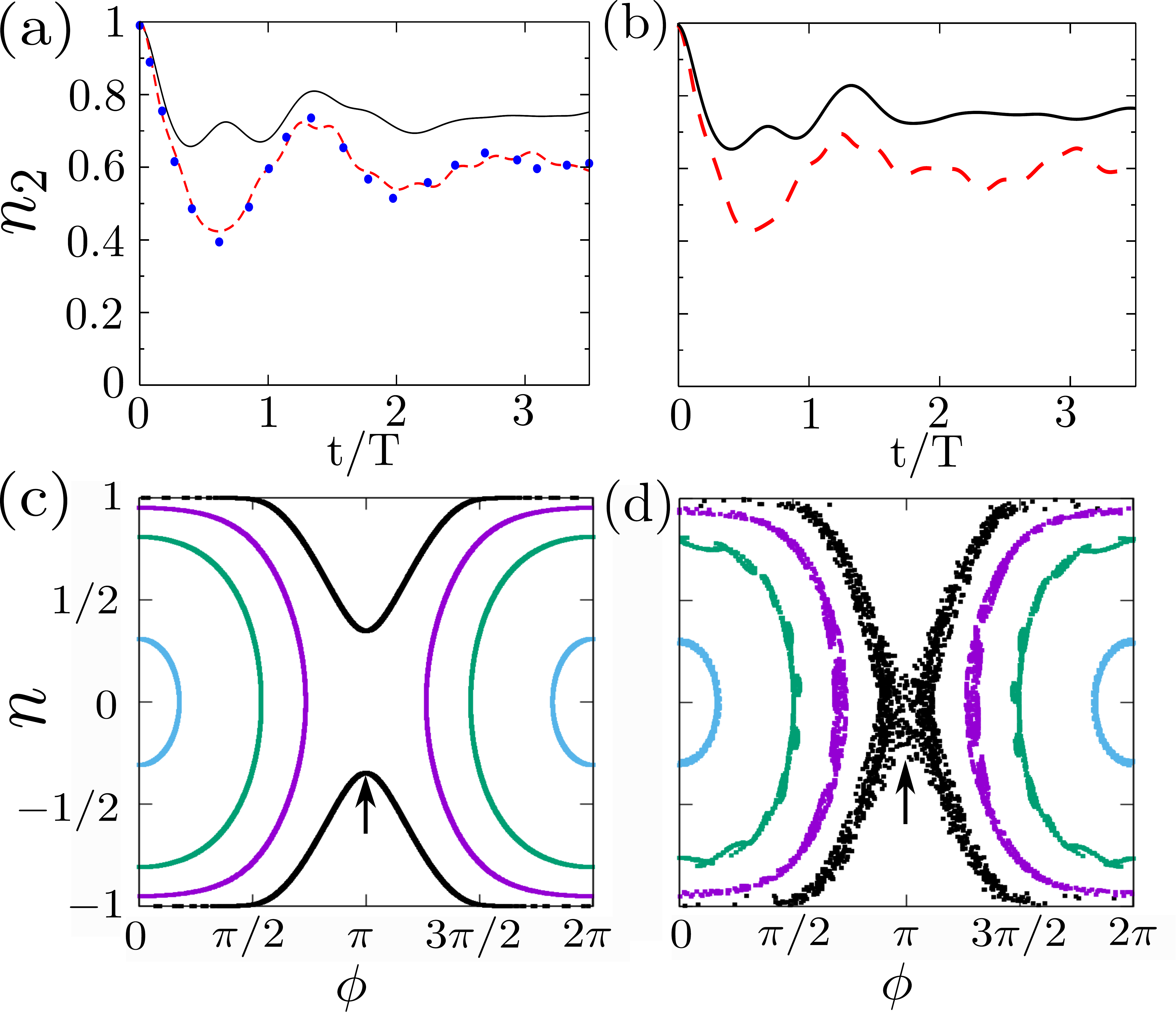}
\caption{(Color online) Comparison with Ref.~\cite{Cederbaum}: (a) Quantum dynamics of $n_2$ according to the BHM (solid black), the MCTDHB (Blue circles), and our simple model (dashed red). MCTDHB data is extracted from Fig.~1d in Ref.~\cite{Cederbaum}; (b) The corresponding semiclassical dynamics with the same color code; (c) Classical phase space structure of the isolated two mode BHM ($\kappa=0$); (d) Corresponding Poincare sections at $\phi_0=0$ in the presence of the excited mode with detuning $\Omega=5$ and coupling $\kappa=0.75$.}
\label{q_dyn}
\end{figure}

\section{Reconstruction of BHM breakdown} 
Returning to the results of Ref.~\cite{Cederbaum}, in \Fig{q_dyn}(a) we compare representative MCTDHB population dynamics with the quantum dynamics of our simple model (\Eq{BHM_trimer}), for the same parameters and initial conditions. The BHM breakdown in \cite{Cederbaum} is clearly captured by the $2+1$ mode model. The quantum results are reproduced by the semiclassical simulation in \Fig{q_dyn}(b), attained by propagating a cloud of classical  trajectories from \Fig{q_dyn}(c)-(d). The observed failure of the BHM is therefore unequivocally related to the persistent separatrix chaos. In fact, as shown in Appendix\ \ref{A2} {\em all} the results of Ref.~\cite{Cederbaum} are reproduced by our model, including the thermalization of a self-trapped trajectory for large values of $u$ where the validity condition \Eq{eStandard} is violated and the two mode approximation breaks down. This thermalization too, is the result of chaotic ergodization of the type shown in \Fig{pp_cl}(c). 


\section{Summary} 
The description of optical lattices by the BHM should not be taken for granted. It is distinct from the truncation of a high-lying band of states in electronic systems that features a spectral gap \cite{Sowinski16}. In any $M$-site BHM the phase space is typically {\em mixed}, meaning that there are possibly vast regions of quasi-regular dynamics. Using the stochastic pump paradigm that leads to Arnold diffusion, we have shown that  the effect of far-detuned modes requires to forego naive reasoning. The effect is amplified due to the many-body mixing of the eigenstates, and provides new insight for the relevance of the formal MCTDHB.
  
\section*{Acknowledgements}
This research was supported by the Israel Science Foundation (Grant  No. 283/18).

\appendix

\section{Estimate of BHM parameters}
\label{A1_def}

The Hamiltonian describing the bosons in a double well trapping potential can be written as \cite{Smerzi97},
\begin{eqnarray}
\hat{H}_{BH} &=& \int dx\, \hat{\psi}^{\dagger}(x)\left(-\frac{\hbar^2}{2m}\partial_x^2 + V(x) \right) \hat{\psi}(x) \nonumber \\
&+& \frac{\lambda_0}{2} \int dx\, \hat{\psi}^{\dagger}(x) \hat{\psi}^{\dagger}(x) \hat{\psi}(x) \hat{\psi}(x) 
\label{BHM1}
\end{eqnarray} 
where, $V(x)$ denotes the double well potential, $\lambda_0$ is the 1D interaction strength and $\hat{\psi}(x)$ is the bosonic field operator. Representing the wave functions, at the lowest quasi-degenerate orbitals (`1' and `2') of the double well by $\Phi_1(x)$ and $\Phi_2(x)$ respectively, and at the excited orbital by $\Phi_0(x)$, we can expand $\hat{\psi}(x)$ as follows:
\begin{equation}
\hat{\psi}(x) = \Phi_1(x)~\hat{a}_1 + \Phi_2(x)~\hat{a}_2 + \Phi_0(x)~\hat{a}_0
\label{boson_op}
\end{equation}
where, $\hat{a}_{1,2,0}$ are the bosonic annihilation operators corresponding to the three modes `1', `2' and `0' respectively. We define the two-mode parameters as in \cite{Smerzi97} and add the coupling parameters to the third mode :\\ 
\begin{equation}
\kappa = -\int \left[\frac{\hbar^2}{2m}(\partial_x\Phi_0 ~ \partial_x\Phi_{1,2}) + \Phi_0 V(x) \Phi_{1,2}\right]dx~,
\end{equation} 

\begin{equation}
\Omega = E_0 - E_{1,2}~, 
\end{equation}
where,
$$
E_{1,2,0} = -\int \left[\frac{\hbar^2}{2m}|\partial_x\Phi_{1,2,0}|^2 + V(x) |\Phi_{1,2,0}|^2\right]dx~.
$$
Substituting Eq.\ \ref{boson_op} into Eq.\ \ref{BHM1} and using the definitions of the above parameters, we obtain the trimer Hamiltonian \eqref{BHM_trimer}, where the single detuned bosonic mode `0' is coupled to two modes (`1' and `2') of the dimer. 

\begin{figure}[t]
\centering
\includegraphics[width =1.\columnwidth]{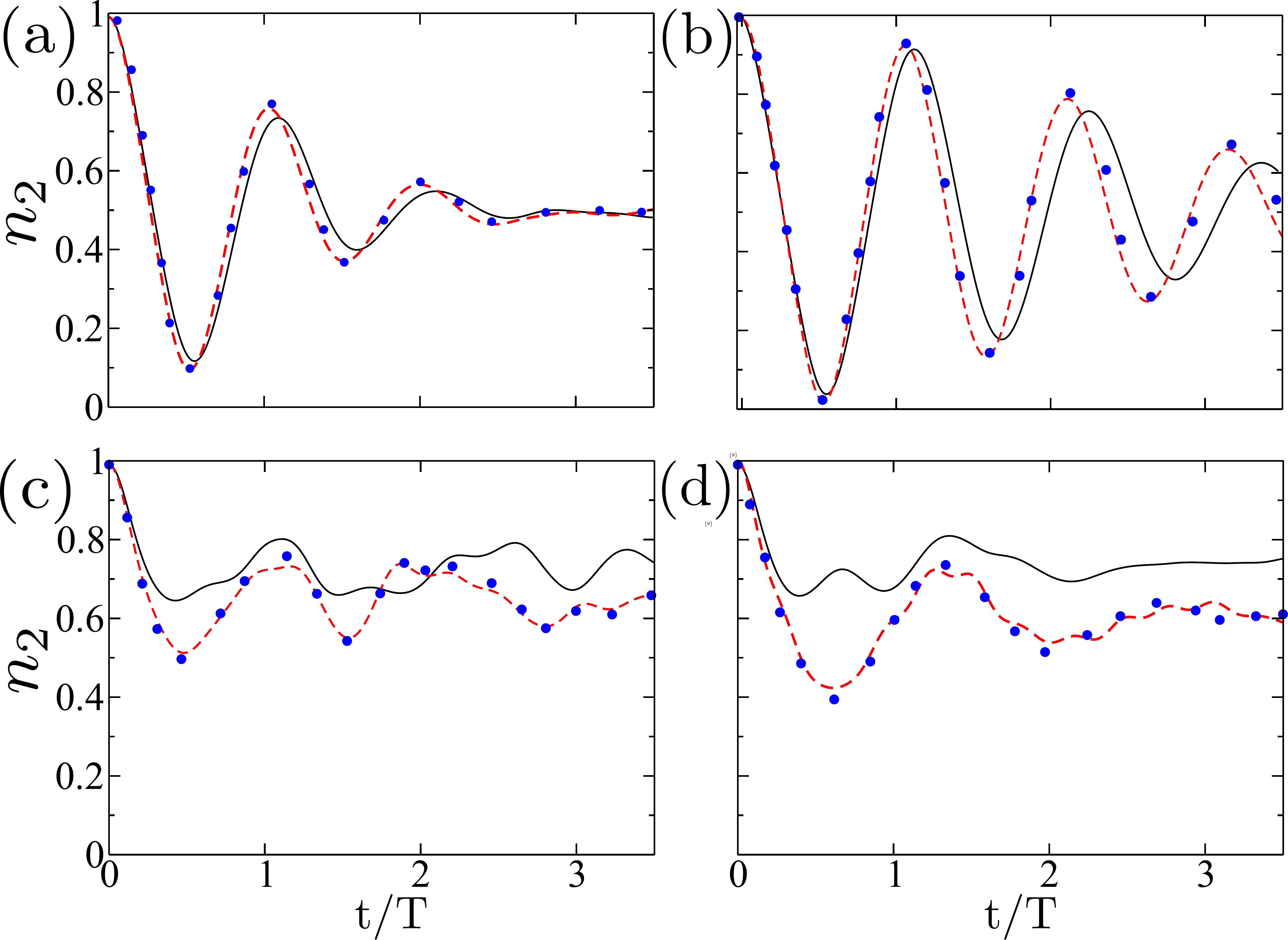}
\caption{(Color online) Breakdown of the BHM despite weak interaction: Comparison of population dynamics  obtained from quantum dynamics using the two-mode BHM (solid black line), the MCTDHB method of Ref.~\cite{Cederbaum} (blue circles), and our $2+1$ mode model \Eq{BHM_trimer} (dashed red line). The parameters and the initial preparation with $n_2=N$ are identical to Fig.~1 of Ref.~\cite{Cederbaum}: (a) $u=1.4$, $N=20$; (b) $u=1.35$, $N=100$; (c) $u=2.26$, $N=20$; (d) $u=2.17$, $N=100$. The parameters used in our model are $\Omega=5$ in all panels, $\kappa=0.65$ in a,b and $0.75$ in c,d.}
\label{fig:S1}
\end{figure} 

\section{Dynamical equations}
\label{A1}

To study the dynamics governed by the Hamiltonian of \Eq{BHM_trimer}, we first write down the Heisenberg equations of motion $i \dot{\hat{a}}_{i} = [\hat{a}_{i},\hat{\mathcal{H}}]$, for the bosonic annihilation operators $\hat{a}_i$ :
\begin{subequations}
\begin{eqnarray}
i\dot{\hat{a}}_1 &=& -\frac{1}{2}\hat{a}_2 + \frac{U}{2}(\hat{n}_1\hat{a}_1 + \hat{a}_1\hat{n}_1) - \frac{\kappa}{2}\hat{a}_0 ~,\label{eqom1} \\
i\dot{\hat{a}}_2 &=& -\frac{1}{2}\hat{a}_1 + \frac{U}{2}(\hat{n}_2\hat{a}_2 + \hat{a}_2\hat{n}_2) - \frac{\kappa}{2}\hat{a}_0 ~,\label{eqom2} \\
i\dot{\hat{a}}_0&=& \Omega \hat{a}_0 - \frac{\kappa}{2}(\hat{a}_1+\hat{a}_2) + \frac{U}{2}(\hat{n}_0\hat{a}_0 + \hat{a}_0\hat{n}_0)~. \label{eqom3}
\end{eqnarray} \label{eqom}
\end{subequations} 
Two conserved quantities constrain the dynamics: One is the total energy and the other is the total number of particle resulting from $[\hat{\mathcal{H}}, \hat{N}]=0$ where, $\hat{N} = \sum_{i=0}^2 \hat{n}_i$. 

\paragraph{Classical dynamics:} In the large $N$-limit the field operators can be replaced by $c$-numbers such as, $\hat{a}_i \rightarrow a_i$, $\hat{n}_{i} \rightarrow n_i$, where, $a_i = \sqrt{n_i}e^{i\phi_i}$, $n_i = |\alpha_i|^2$.
The classical populations and phases $\{n_{i}, \phi_{i}\}$ serve as conjugate dynamical variables. The equation of motion for the complex variables $a_i$ are thus given by,
\begin{subequations}
\begin{eqnarray}
i\dot{a}_1 &=& -\frac{1}{2}a_2 + u|a_1|^2 a_1 - \frac{\kappa}{2}a_0 \label{eqom1_cl} ~,\\
i\dot{a}_2 &=& -\frac{1}{2}a_1 + u|a_2|^2 a_2 - \frac{\kappa}{2}a_0 \label{eqom2_cl} ~,\\
i\dot{a}_0 &=& \Omega a_0 - \frac{\kappa}{2}(a_1+a_2) + u|a_0|^2 a_0~, 
\label{eqom3_cl}
\end{eqnarray}
\label{eqom_cl}
\end{subequations}
where, $u = UN/K$. Using $N$ conservation, we eliminate one degree of freedom, leaving $n = n_1 - n_2$, $\phi = \phi_1 - \phi_2$, $n_0$, and $\phi_0$ as the dynamical variables of our two-freedoms system. Evolving \Eq{eqom_cl}, we obtain the classical results of \Fig{pp_cl}.

\begin{figure}[t]
\centering
\includegraphics[width =1.\columnwidth]{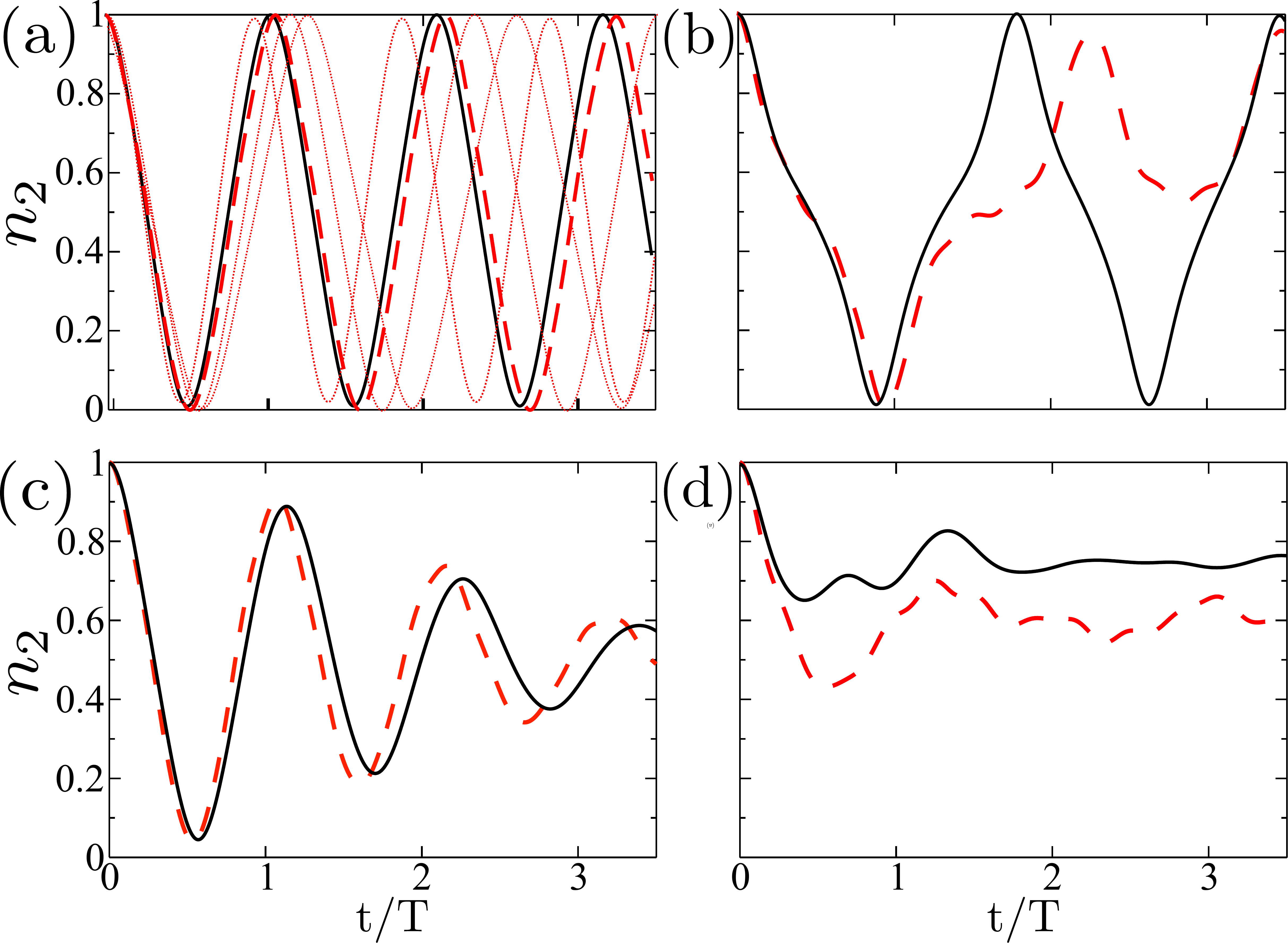}
\caption{(Color online) Semiclassical dynamics: (a) Classical Gross-Pitaevskii simulations, corresponding to \Fig{fig:S1}b with the same color code. Representative trajectories launched in the vicinity  of the classical initial conditions are shown as thin dotted lines; (b) Same for \Fig{fig:S1}d, with no additional trajectories; (c) Semiclassical dynamics, obtained from the classical propagation of an initially Gaussian cloud with width $\sqrt{2/N}$, reproduce the quantum results of \Fig{fig:S1}b; (d) Similarly, semiclassical propagation reproduces  \Fig{fig:S1}d}.
\label{fig:S1a}
\end{figure}


\paragraph{Quantum dynamics:} The quantum Fock states of the three-mode models are $|n, n_0\rangle$ where, $n_0 \in [0,N]$ and $n \in [-(N-n_0),(N-n_0)]$. The operation of the field operators is e.g. $\hat{a}_1^{\dagger} \hat{a}_2 \, | n, n_0\rangle = \sqrt{n_2(n_1+1)} \, |n+2, n_0\rangle$ where, $n_1 = (N-n_0+n)/2$ and $n_2 = (N-n_0-n)/2$. Given some initial quantum preparation $|\Psi(0)\rangle$, we employ the Fock state representation to obtain the time evolved state $|\Psi(t)\rangle = e^{-i \hat{\mathcal{H}} t} |\Psi(0)\rangle$.
In comparing quantum and classical dynamics, the initial quantum states $|\Psi(0)\rangle$ are three-mode coherent states,  
\begin{equation}
|n,\phi,n_0,\phi_0\rangle =  \frac{1}{\sqrt{N!}} 
\left( \sum_{i=0}^2 \alpha_i \hat{a}_i^{\dagger} \right)^N \, |\text{Vac}\rangle~, 
\end{equation} 
where $\alpha_1=\sqrt{(N+n)/2}$ and $\alpha_2=\sqrt{(N-n)/2}e^{i\phi}$ are the same as in the classical simulations and $\alpha_0=\sqrt{n_0}e^{i\phi_0}$ is set to zero.

\paragraph{Semiclassical dynamics:}
The classical dynamics can only capture the quantum evolution up to the Ehrenfest time. However, averaging over many classical trajectories offers accuracy over longer timescales. In semiclassical simulations, we prepare a cloud of classical points that emulate the initial quantum phase-space distribution. Initial coherent states correspond to minimal Gaussian clouds with $2/N$ variance. Each point is then propagated classically and the values of observables are obtained by averaging over the cloud. This procedure is essentially a truncated Wigner phase-space approach.

\begin{figure}[t]
\centering
\includegraphics[width =1.\columnwidth]{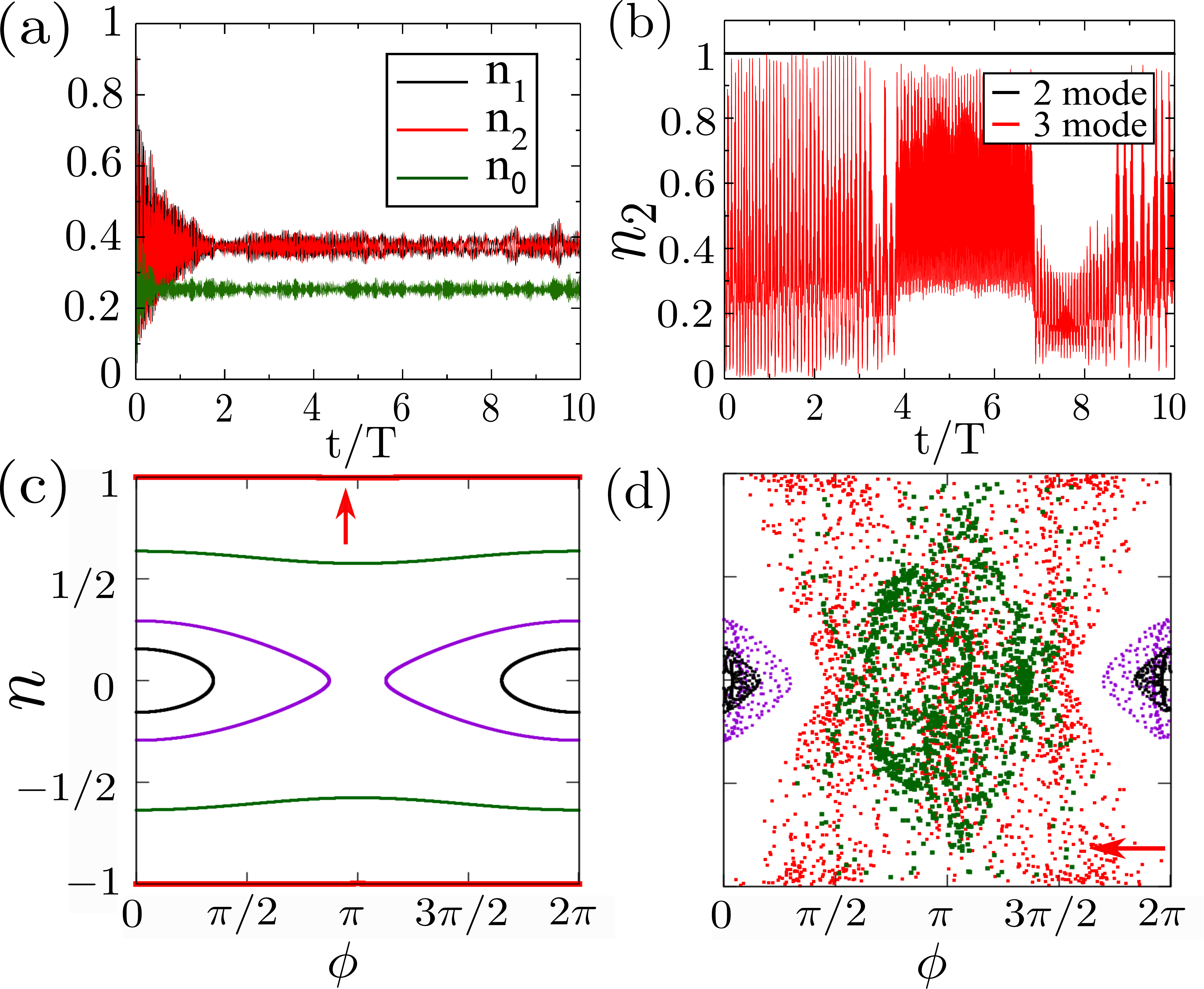}
\caption{(Color online) Chaotic ergodization at strong interaction: (a) Quantum population dynamics obtained by time propagation with the Hamiltonian of \Eq{BHM_trimer} with $\Omega=30$ and $\kappa=40$, for the same parameters as in Fig. 2 of Ref.~\cite{Cederbaum}: $u=43.4$, $N=100$; (b) Classical population dynamics for the same parameters using either the two-mode BHM (black line,  depicting a nearly-stationary point at $n_2=N$) or the Hamiltonian of \Eq{BHM_trimer} (red line, exploring the entire allowed population range); (c) The two-mode phase-space; (d) Poincare sections in the $2+1$ mode phase space, as in \Fig{pp_cl}. The self-trapped dimer at $n_2=N$ becomes chaotic due to the coupling to the auxiliary mode and explores the entire stochastic band, resulting in the 'thermalization' of the population distribution.}
\label{fig:S2}
\end{figure} 


\section{Comparison with previous numerical results} 
\label{A2}

In Ref.~\cite{Cederbaum}, deviations from the two-mode BHM are observed numerically, by employing the multiconfigurational time-dependent Hartree for bosons (MCTDHB) method, to obtain exact double-well dynamics that include the effect of excited modes. The dynamics shown in Fig.~1 of  \cite{Cederbaum}, that is reproduced here in \Fig{fig:S1}, demonstrate that the BHM breaks down even when $u$ is small with respect to $\Omega/K$ and \Eq{eStandard} is satisfied. The initial preparation with $n_2=N$ gives Rabi-Josephson oscillations for $u<2$ or self-trapped motion for $u>2$. For $u\approx 2$ it lies near the separatrix. Thus, Panels (a) and (b) illustrate the validity of the BHM for Rabi-Josephson oscillations, but panels (c) and (d) show its failure to depict near-separatrix motion.

The dashed red lines in \Fig{fig:S1} corresponds to quantum propagation using the Hamiltonian of \Eq{BHM_trimer}. Our simple model clearly reproduces the results of the MCTDHB model (blue circles) to great accuracy. The relation between the quantum failure of the BHM and the classical persistence of chaos in the vicinity of the separatrix, is made clear by inspection of the semiclassical simulations presented in  \Fig{fig:S1a}. The dynamical instability of the classical trajectories that constitute the semiclassical cloud leads to strong deviations from the BHM in the vicinity of the separatrix, thereby reproducing the observed quantum result. We therefore deduce that the numerical observations in Ref.~\cite{Cederbaum} can be attributed in essence, to our mechanism of BHM failure due to chaos.

The observed agreement extends beyond the weak-interaction regime where the BHM validity condition is satisfied. In Fig.~2 of Ref.~\cite{Cederbaum}, the interaction strength is large enough to violate \Eq{eStandard}. Thus, while the $n_2=N$ preparation is deep in the self trapped region in this case, so that almost no population oscillation exists in its two-mode dynamics, the excited mode can affect the entire phase-space, and in particular can transform self-trapped trajectories to chaotic ones, as in \Fig{pp_cl}(c).  The result is the thermalization of the population distribution that was noted in \cite{Cederbaum}. In \Fig{fig:S2} we reproduce this result for the pertinent parameters, showing that it too is attributed to chaos. 



\begin{thebibliography}{99}

\bibitem{Fisher89}
M. P. A. Fisher, P. B. Weichman, G. Grinstein, and D. S. Fisher,
Phys. Rev. B {\bf 40}, 546 (1989).

\bibitem{Leggett01}
A.J. Leggett, Rev. Mod. Phys. {\bf 73}, 307 (2001).

\bibitem{Hubbard63}
J. Hubbard, Proc. R. Soc. Lond. Ser. A Math. Phys. Eng. Sci., {\bf 276}, 238 (1963). 

\bibitem{Gutzwiller63}
M.C. Gutzwiller 
Phys. Rev. Lett. {\bf 10}, 159 (1963). 

\bibitem{Bloch05}
I. Bloch,
Nat. Phys. {\bf 1}, 23 (2005).

\bibitem{Krutitsky16}
K. V. Krutitsky,
Phys. Rep. {\bf 607},  1 (2016).

\bibitem{Jaksch98}
D. Jaksch, C. Bruder, J. I. Cirac, C. W. Gardiner, P. Zoller, 
Phys. Rev. Lett. {\bf 81}, 3108 (1998).

\bibitem{Orzel01}
C. Orzel, A. K. Tuchman, M. L. Fenselau, M. Yasuda, M. A. Kasevich, 
Science {\bf 291}, 2386 (2001).

\bibitem{Greiner02a}
M. Greiner, O. Mandel, T. Esslinger, T. W. H\"ansch, I. Bloch,
Nature {\bf 415}, 39 (2002).

\bibitem{Greiner02b}
M. Greiner, O. Mandel, T. W. H\"ansch, I. Bloch, 
Nature {\bf 419}, 51 (2002).

\bibitem{Will10}
S. Will, T. Best, U. Schneider, L. Hackerm\"uller, D. S. R. L\"uhmann, and I. Bloch,
Nature {\bf 465}, 197 (2010).

\bibitem{Lewenstein07}
M. Lewenstein, A. Sanpera, V. Ahufinger, B. Damski,  A. Sen(De), and U. Sen,
Adv. Phys., {\bf 56}, 243 (2007).

\bibitem{Bloch12}
I. Bloch, J. Dalibard, and S. Nascimbene,
Nat. Phys. {\bf 8}, 267 (2012).

\bibitem{LMG}
H. J. Lipkin, N. Meshkov, and A. J. Glick, Nucl. Phys. {\bf 62}, 188 (1965);  N. Meshkov, A. J. Glick, and H. J. Lipkin, Nucl. Phys. {\bf 62}, 199 (1965); A. J. Glick, H. J. Lipkin, and N. Meshkov, Nucl. Phys. {\bf 62}, 211 (1965).

\bibitem{Jo07}
G.-B. Jo, Y. Shin, S. Will, T. A. Pasquini, M. Saba, W. Ketterle, D. E. Pritchard, M. Vengalattore and M. Prentiss,
Phys. Rev. Lett. {\bf 98}, 030407 (2007).

\bibitem{Esteve08}
J. Esteve, C. Gross, A. Weller, S. Giovanazzi, and M. K. Oberthaler,
Nature {\bf 455}, 1216 (2008).

\bibitem{Riedel10}
M. F. Riedel, P.l B\"ohi, Yun Li, T. W. H\"ansch, A. Sinatra, and P. Treutlein
Nature {\bf 464}, 1170 (2010).

\bibitem{Milburn97}
G. J. Milburn, J. Corney, E. M. Wright, and D. F. Walls
Phys. Rev. A {\bf 55}, 4318 (1997).

\bibitem{Smerzi97}
A. Smerzi, S. Fantoni, S. Giovanazzi, and S. R. Shenoy,
Phys. Rev. Lett. {\bf 79}, 4950 (1997).

\bibitem{Albiez05}
M. Albiez, R. Gati, J. F\"olling, S. Hunsmann, M. Cristiani, and M. K. Oberthaler
Phys. Rev. Lett. {\bf 95}, 010402 (2005).

\bibitem{Gati07}
R. Gati, M.K. Oberthaler, J. Phys. B {\bf 40}, 61(R) (2007).

\bibitem{Levy07}
S. Levy, E. Lahoud, I. Shomroni, and J. Steinhauer,
Nature {\bf 449}, 579 (2007).

\bibitem{Vardi01}
A. Vardi and J. R. Anglin, 
Phys. Rev. Lett. {\bf 86}, 568 (2001).

\bibitem{Anglin01}
J. R. Anglin and A. Vardi,
Phys. Rev. A {\bf  64}, 013605 (2001).

\bibitem{Khodorkovsky08}
Y. Khodorkovsky, G. Kurizki, and A. Vardi, 
Phys. Rev. Lett. {\bf 100}, 220403 (2008).

\bibitem{Boukobza09}
E. Boukobza, M. Chuchem, D. Cohen and A. Vardi, 
Phys. Rev. Lett. {\bf 102}, 180403 (2009).

\bibitem{Boukobza10}
E. Boukobza, M. G. Moore, D. Cohen, and A. Vardi, 
Phys. Rev. Lett. {\bf 104}, 240402 (2010).

\bibitem{Chuchem10}
M. Chuchem, K. Smith-Mannschott, M. Hiller, T. Kottos, A. Vardi, and D. Cohen, 
Phys. Rev. A, 82, 053617 (2010).

\bibitem{Wright13}
K. C. Wright, R. B. Blakestad, C. J. Lobb, W. D. Phillips, and G. K. Campbell,
Phys. Rev. Lett. {\bf 110}, 025302 (2013).

\bibitem{Eckel14}
S. Eckel, J. G. Lee, F. Jendrzejewski, N. Murray, C. W. Clark, C. J. Lobb, W. D. Phillips, M. Edwards, and G. K. Campbell,
Nature {\bf 506}, 200 (2014).

\bibitem{Fialko12}
O. Fialko, M.-C. Delattre, J. Brand, and A. R. Kolovsky,
Phys. Rev. Lett. {\bf 108}, 250402 (2012).

\bibitem{Arwas16}
G. Arwas and D. Cohen,,
New J. Phys. {\bf 18}, 015007 (2016).

\bibitem{Arwas17a}
G. Arwas and D. Cohen,
Phys. Rev. B {\bf 95}, 054505 (2017).

\bibitem{Arwas17b}
Geva Arwas, Doron Cohen, Frank Hekking, and Anna Minguzzi,
Phys. Rev. A {\bf 96}, 063616 (2017).

\bibitem{Cederbaum} K. Sakmann, A. I. Streltsov, O. E. Alon and L. S. Cederbaum, Phys. Rev. Lett. {\bf 103}, 220601 (2009).

\bibitem{Bergeman06} D. Ananikian and T. Bergeman, Phys. Rev. A {\bf 73}, 013604 (2006).

\bibitem{Trujillo-Martinez09}
M. Trujillo-Martinez, A. Posazhennikova, and J. Kroha,
Phys. Rev. Lett. {\bf 103}, 105302 (2009).

\bibitem{Smerzi11} L. J. LeBlanc, A. B. Bardon, J. McKeever, M. H. T. Extavour, D. Jervis, J. H. Thywissen, F. Piazza, and A. Smerzi, Phys. Rev. Lett. {\bf 106}, 025302 (2011).

\bibitem{Modugno18} Sof\'ia Mart\'inez-Garaot, Giulio Pettini, and Michele Modugno,
Phys. Rev. A {\bf 98}, 043624 (2018).

\bibitem{Cataldo13} D. M. Jezek, P. Capuzzi, and H. M. Cataldo, Phys. Rev. A {\bf 87}, 053625 (2013).

\bibitem{Chirikov}
Boris. V. Chirikov, Phys. Rep. {\bf 52}, 263 (1979).

\bibitem{Guan03} H.Q. Zhou, J. Links, R.H. McKenzie, and X.W. Guan, J. Phys. A {\bf 36}, L113 (2003).

\bibitem{Korsch09} F. Trimborn, D. Witthaut, and H. J. Korsch, Phys. Rev. A {\bf 79}, 013608 (2009).

\bibitem{Hiller09} M. Hiller, T. Kottos, and T. Geisel,
Phys. Rev. A {\bf 79}, 023621 (2009).

\bibitem{Liu07} Bin Liu, Li-Bin Fu, Shi-Ping Yang, and Jie Liu, Phys. Rev. A {\bf 75}, 033601 (2007).

\bibitem{Jung06} S. Mossmann and C. Jung, Phys. Rev. A {\bf 74}, 033601 (2006).

\bibitem{Penna03} P. Buonsante, R. Franzosi, and V. Penna, Phys. Rev. Lett. {\bf 90}, 050404 (2003).

\bibitem{Tikhonenkov13} I Tikhonenkov, A. Vardi, J. R. Anglin, and D. Cohen, 
Phys. Rev. Lett. {\bf 110}, 050401 (2013).

\bibitem{Gallemi15} A Gallemí, M Guilleumas, J Martorell, R Mayol, A. Polls and B. Juli\'a-Díaz, 
New J. Phys. {\bf 17}, 073014 (2015).

\bibitem{Dey18} A. Dey, D. Cohen, and A. Vardi, 
Phys. Rev. Lett. {\bf 121}, 250405 (2018).

\bibitem{Burkle19}
R. B\"urkle, A. Vardi, D. Cohen, and J. R. Anglin, 
Phys. Rev. Lett. {\bf 123}, 114101 (2019).

\bibitem{Flambaum}
V. V. Flambaum,
Phys. Scr. {\bf 1993}, 198 (1993).

\bibitem{Sowinski16}
Even few-particle systems impose severe limitations, see J. Dobrzyniecki and T. Sowi\'nski, Eur. Phys. J. D {\bf 70}, 83 (2016).



\end{thebibliography}
\end{document}